\documentclass{aa}  

\usepackage{graphicx}
\usepackage{txfonts}
\usepackage{color}
\usepackage{booktabs}

\definecolor{linka}{rgb}{0.0,0.0,0.7}
\definecolor{linkb}{rgb}{0.0,0.0,0.5}
\definecolor{linkc}{rgb}{0.0,0.0,0.3}
\definecolor{cmt}{rgb}{0.0,0.4,0.0}
\definecolor{al}{rgb}{0.6,0.2,0.0}
\definecolor{mvn}{rgb}{0.4,0.4,0.0}
\definecolor{referee}{rgb}{0.4,0.0,0.0}
\usepackage[colorlinks=true,linkcolor=linkb,citecolor=linkc,filecolor=linka,urlcolor=linka,bookmarks=true,bookmarksopen=true,bookmarksopenlevel=3,plainpages=false,pdfpagelabels=true]{hyperref}
\usepackage{natbib}

\newcommand{\kms}{km\,s$^{-1}$}
\newcommand{\ms}{m\,s$^{-1}$}
\newcommand{\grad}{$^\circ$}
\newcommand{\carcsec}{$\mbox{.\hspace{-0.5ex}}^{\prime\prime}$}

\newcommand{\fei}{Fe\,\textsc{i}}

\newcommand{\caiih}{Ca\,\textsc{ii\,h}}

\newcommand{\los}{${\rm LOS}$}
\newcommand{\vlos}{$v_{\rm LOS}$}
\newcommand{\vmici}{$v_{\rm mic}$}

\newcommand{\lbg}{LBG}
\newcommand{\plg}{P\textsc{l}G}
\newcommand{\qsg}{QSG}

\newcommand{\fig}[1]{Fig.~\ref{#1}} 
\newcommand{\eq}[1]{Eq.~\ref{#1}} 
\newcommand{\figs}[1]{Figs.~\ref{#1}} 
\newcommand{\tab}[1]{Tab.~\ref{#1}} 
\newcommand{\sect}[1]{Sec.~\ref{#1}} 
\newcommand{\secs}[1]{Sections \ref{#1}} 
 
\newcommand{\oapp}[1]{online App.~\ref{#1}} 

\newcommand{\bx}[1]{\textsl{(B#1)}} 
\newcommand{\cut}[1]{\textsl{(C#1)}} 
\def \figpath {./}

\newcommand{\colfigtwocol}[3][1.]{\begin{figure*}\centering
    \includegraphics[width=#1\linewidth,clip=TRUE]{\figpath#2.eps}
    \caption{#3}
    \label{#2}
\end{figure*}}
\newcommand{\colfig}[3][1.]{\begin{figure}\centering
    \includegraphics[width=#1\linewidth,clip=TRUE]{\figpath#2.eps}
    \caption{#3}
    \label{#2}
\end{figure}}

\begin{document}

\title{Vigorous convection in a sunspot granular light bridge}


\author{Andreas Lagg
  \inst{1}
  \and
  Sami K. Solanki\inst{1,2}
  \and
  Michiel van Noort\inst{1}
  \and
  Sanja Danilovic\inst{1}
}

\institute{Max Planck Institute for Solar System Research, Justus-von-Liebig-Weg 3, 37077 G\"ottingen, Germany
  \and
School of Space Research,
Kyung Hee University, Yongin, Gyeonggi 446-701, Republic of Korea\\
\email{[lagg;solanki;vannoort;danilovic]@mps.mpg.de}}
\offprints{A. Lagg, \email{lagg@mps.mpg.de}}
\date{Received 25/04/2014; accepted 20/06/2014}

\abstract
{Light bridges are the most prominent manifestation of convection in sunspots. The brightest representatives are granular light bridges composed of features that appear to be similar to granules.}
{An in-depth study of the convective motions, temperature stratification, and magnetic field vector in and around light bridge granules is presented with the aim of identifying similarities and differences to typical quiet-Sun granules.}
{Spectropolarimetric data from the Hinode Solar Optical Telescope were analyzed using a spatially coupled inversion technique to retrieve the stratified atmospheric parameters of light bridge and quiet-Sun granules.}
{Central hot upflows surrounded by cooler fast downflows reaching 10~\kms{} clearly establish the convective nature of the light bridge granules. The inner part of these granules in the near surface layers is field free and is covered by a cusp-like magnetic field configuration. We observe hints of field reversals at the location of the fast downflows. The quiet-Sun granules in the vicinity of the sunspot are covered by a low-lying canopy field extending radially outward from the spot.}
{The similarities between quiet-Sun and light bridge granules point to the deep anchoring of granular light bridges in the underlying convection zone. The fast, supersonic downflows are most likely a result of a combination of invigorated convection in the light bridge granule due to radiative cooling into the neighboring umbra and the fact that we sample deeper layers, since the downflows are immediately adjacent to the slanted walls of the Wilson depression.}


\maketitle

\section{Introduction}

The strong magnetic field in sunspots effectively hinders the overturning motion of the plasma in the sunspot umbral photosphere \cite[]{biermann:41,gough:66}. The reduced energy input from below leads to a lowering of the temperature in photospheric layers. This low temperature in combination with the decreased gas density, caused by the displacement of gas due to the magnetic pressure, significantly decreases the opacity of the solar atmosphere, allowing us to see a few hundred kilometers deeper than in quiet regions of the Sun \cite[Wilson depression, ][]{loughhead:58}.

Bright structures within the umbra are signatures of not completely suppressed convection. The best known representatives of these structures are light bridges and umbral dots. Here we concentrate on light bridges.

Light bridges (LBs) can be categorized by their brightness and size. Faint light bridges \cite[FLBs, e.g.,][]{sobotka:09,sobotka:93,lites:91} are elongated structures in sunspot umbrae composed of grains of similar size and structure to umbral dots. Strong light bridges \cite[e.g.,][]{sobotka:93,rimmele:08,rezaei:12} with a typical brightness comparable to the penumbra often separate the umbra into two regions of the same polarity. Granular light bridges (GLBs), sometimes also called photospheric light bridges \cite[e.g.,][]{vazquez:73,lites:91,sobotka:94,leka:97,rouppe:10}, show fully developed convective cells similar to the granules in the quiet-Sun. All types exhibit a significantly reduced field strength in the photosphere with a cusp-like magnetic field configuration at higher layers \cite[]{jurcak:06}, which for faint and strong light bridges is accompanied by a central dark lane originating from an enhanced density in the cusp \cite[]{spruit:10a}. Central upflows surrounded by downflows toward the umbra, sometimes at supersonic speeds \cite[]{louis:09,bharti:13a}, point to the convective origin of LBs.
Such high-speed downflows at the edges of granular cells directly adjacent to sunspot umbrae have already been observed by \cite{shimizu:08b}.
The chromospheric activity above light bridges is often enhanced and manifests itself in the form of jets, surges, and brightenings in, say, \caiih{} \cite[]{shimizu:09}. Apart from these dynamic events, the magnetic field configuration in the upper chromospheric layers becomes very similar to the umbral environment \cite[]{ruedi:95a,joshi:14a}.

The mechanism producing granular light bridges is believed to be distinctively different from the formation of other convective phenomena in sunspots like penumbral filaments or umbral dots. Whereas the latter are believed to be the consequence of magneto-convection within a 1--2~Mm thick layer beneath the photosphere \cite[]{schuessler:06,rempel:09a,rempel:09b}, thicker light bridges are attributed to intrusions of field-free plasma from deep beneath sunspots \cite[]{rempel:11a,leka:97} or to the inward motion of hot gas from the penumbra triggered by sub-photospheric flows crossing the sunspot \cite[]{katsukawa:07a}. Broad light bridges often consist of several granular convection cells along the light bridge axis. The presence of such granulation cells embedded in the low-density, transparent environment of sunspot umbrae and the resulting exposure of their walls allow probing the physical conditions in deep layers of such cells, which are otherwise not accessible to direct observation.

This possibility motivated us to investigate the properties of granular light bridges based on the physical parameters determined from spatially coupled inversions of Hinode spectro-polarimetric data. The observations and the analysis method are described in \secs{obs} and \ref{method}, respectively. We present the properties of light bridge granules in \sect{result} and compare these with their quiet-Sun counterparts. We then discuss the observed configuration in \sect{discussion} and summarize the results in \sect{conclusion}.

\section{Observations\label{obs}}

Spectropolarimetric data of the leading spot of AR10926 were obtained using the spectropolarimeter of the Hinode Solar Optical Telescope \cite[SOT/SP,][]{kosugi:07a,tsuneta:08a,suematsu:08a,ichimoto:08a,shimizu:08a,lites:13a} on November 30 2006 UT~23:40--00:05. The center of the analyzed region was located at the solar position $x=-184$\arcsec{}, $y=-160$\arcsec{}, corresponding to a heliocentric angle of $\Theta=14.5^\circ{}$ ($\mu=\cos\Theta=0.97$). SOT/SP was operated in ``normal map'' mode with a pixel size of 0\carcsec{}16 in both, the slit and the scan direction. The exposure time per slit was 4.8~s, resulting in a noise level in the quiet-Sun of typically $1.1\times 10^{-3}~I_C$ in Stokes $Q$, $U$, and $V$. Standard data reduction tools from SolarSoft were applied to reduce the data \cite[]{lites:13b}. Averages over the dark umbral regions of the sunspot were used to determine the central wavelength of the \fei{} lines, defining the averaged line-of-sight (\los{}) velocity in the umbra to be zero.

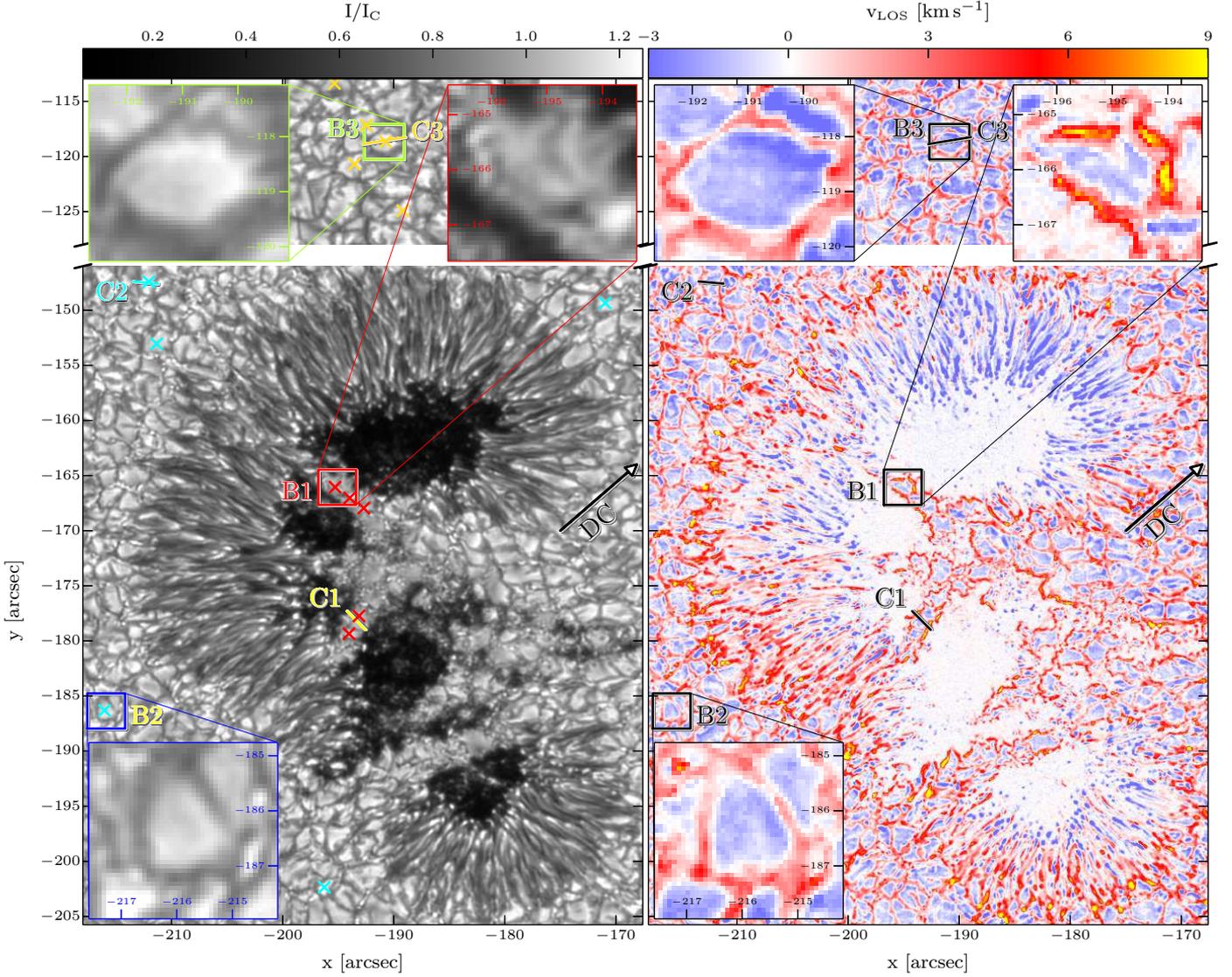
\colfigtwocol{icont_vlos}{Continuum map of AR10926 composed from the fitted Stokes $I$ profiles (left) and \los{} velocity map at $\log\tau=0.0$ (right). Positive values (red/yellow colors) denote downflows. The direction toward the disk center (DC) is indicated by the black arrow. The boxes \bx{1}, \bx{2}, and \bx{3}, and the lines \cut{1} (in the center), \cut{2} (top left), and \cut{3} (top center) correspond to regions analyzed in \sect{result}. The red, cyan, and yellow crosses in the continuum image mark the granules used to determine the average atmospheric parameters in \tab{avgpar} for \lbg{}s, \plg{}s, and \qsg{}s, respectively.}

The lefthand panel of \fig{icont_vlos} displays the continuum image of AR10926 composed from the individual fits to the Stokes $I$ profiles of the SOT/SP observation on the interpolated grid (pixel size 0\carcsec{}08, see \sect{method}). AR10926 came across the east limb on November 24 2006. SOHO/MDI images indicate that already then the sunspot umbra was divided into several unipolar regions separated by strong light bridges. The decaying sunspot disappeared behind the west limb on December 8. The segmentation of the umbra into 3--6 parts separated by strong, granular light bridges could be observed from November 26 until December 4, indicating that the sunspot structure presented in \fig{icont_vlos} represents a quite stable configuration (see the animation attached to \fig{gband_gap}).
The movie of the temporal evolution of the sunspot obtained using the Hinode SOT Broad-band Filter Imager (BFI) at the time of the SOT/SP scan in G-band (left) and \caiih{} (right) is attached to \fig{ca_gb}\footnote{All movies are also available on the website of the Max Planck Institute for Solar System Research: \\
\href{http://www.mps.mpg.de/homes/lagg/OnlineMaterial/2014_LightBridge/}{www.mps.mpg.de/homes/lagg/OnlineMaterial/2014\_LightBridge/} \\
 G-band (long time series): \href{http://www.mps.mpg.de/homes/lagg/OnlineMaterial/2014_LightBridge/gband_gap.avi}{gband\_gap.avi}, G-band and \caiih{} (short time series): \href{http://www.mps.mpg.de/homes/lagg/OnlineMaterial/2014_LightBridge/ca_gb.avi}{ca\_gb.avi}}. These movies indicate that there was no enhanced activity, beyond what is normal for the immediate  surroundings of sunspots, visible above the light bridges, neither in the G-band nor in the \caiih{} line. Note the dark lines running roughly along the middle of some of the light bridges in the \caiih{] movie. The \los{} velocity map, resulting from the inversion described in \sect{method} (right panel), nicely illustrates the strong downflows all along the edges of the granular light bridges.

The inset \bx{1} in \fig{icont_vlos} shows a typical ``granular'' cell within a granular light bridge. We name these cells ``light bridge granules'' (\lbg{}s). Several dark lanes can be identified in the continuum image indicating a lower temperature at the $\tau=1$ layer.
In this study, we compare the properties of several such \lbg{}s with those of ``plage granules'' (\plg{}s) and ``quiet-Sun granules'' (\qsg{}s) found in the same data set, typical examples of those are shown enlarged in boxes \bx{2} and \bx{3} of \fig{icont_vlos}, respectively. The analyzed \plg{}s are located approximately 5--10\arcsec{} away from the visible penumbra boundary, the region for the \qsg{}s lies approximately 20--30\arcsec{} north of the penumbral boundary (note the break in the $y$-axis in \fig{icont_vlos}), where the weakly polarized Stokes profiles indicate an area of very low magnetic field.

\section{Analysis method\label{method}}

The spectropolarimetric data in the \fei{} 6301.5$/$6302.5~\AA{} line pair were inverted using the SPINOR inversion code \cite[]{frutiger:00a,frutiger:thesis}, which employs the STOPRO routines of \cite{solanki:87}, in its spatially coupled version \cite[]{vannoort:12}. On a grid with a pixel size of 0\carcsec{}08, corresponding to half the SOT/SP pixel and scan step size, this inversion technique self-consistently takes into account the spreading of the photons due to the point-spread-function (PSF) of the telescope \cite[]{vannoort:13a}. This approach was found to reproduce complex multi-lobed profiles with a simple, one-component atmosphere per pixel. The complexity of the profiles appears to be produced mainly by the influence of the PSF. The spatially coupled approach was found to significantly enhance the reliability and the robustness of the inversion results.

Every pixel was fitted using a height dependent atmosphere under local thermodynamic equilibrium (LTE) conditions at three nodes in $\log\tau$, the logarithm to the base 10 of the optical depth. The nodes were placed $\log\tau=-2.0$, $-0.8$, and $0.0$, with the free parameters temperature ($T$), magnetic field strength ($B$), magnetic field inclination with respect to the line-of-sight (\los{}) and azimuth ($\gamma$, $\chi$), \los{} velocity (\vlos{}), and a micro-turbulent velocity (\vmici{}). No further broadening mechanism (e.g., macro-turbulence) or straylight correction were applied.

It is an intrinsic problem of inversions to compute reliable error estimates for the retrieved parameters. The coupling between the pixels through the PSF adds additional complexity to this problem. However, the inversion technique and the model atmosphere applied here were established from extensive testing and comparisons with traditional inversion techniques \cite[]{vannoort:12,buehler:13,riethmueller:13a,tiwari:13a}, and proven to be superior in both, the reliability of the parameter retrieval and the robustness of the convergence. Due to the conceptual difficulty to obtain error bars of spatially coupled inversion results \cite[see section 3 of][]{vannoort:13a} we refrain from quantifying errors for the individual pixels of the inverted maps. However, we would like to point out that the features discussed in detail in the following sections are seen in all granules investigated, irrespective of their position within the sunspot, their orientation, their size, and their surroundings. It appears unlikely to us that a systematic error or a lack of signal would produce a unique solution under such varying conditions.


\section{Atmospheric parameters of granules in light bridges and outside the sunspot\label{result}}

\subsection{Maps of atmospheric parameters\label{maps}}

The temporal evolution of AR10926 (see online material) indicates the convective nature of its light bridges. The long axis of each light bridge consists of a chain of convective cells with widths typical of quiet-Sun granules (1--2\arcsec{}). The atmospheric parameters derived from the inversions of the Stokes parameters allow us to confirm and to characterize the convective nature of these cells. As an example, in \fig{lb_granule.noazi} we present maps of the temperature, \los{} velocity, and the strength and direction of the magnetic field for the three $\log\tau$ nodes used during the inversion of a typical \lbg{}. 

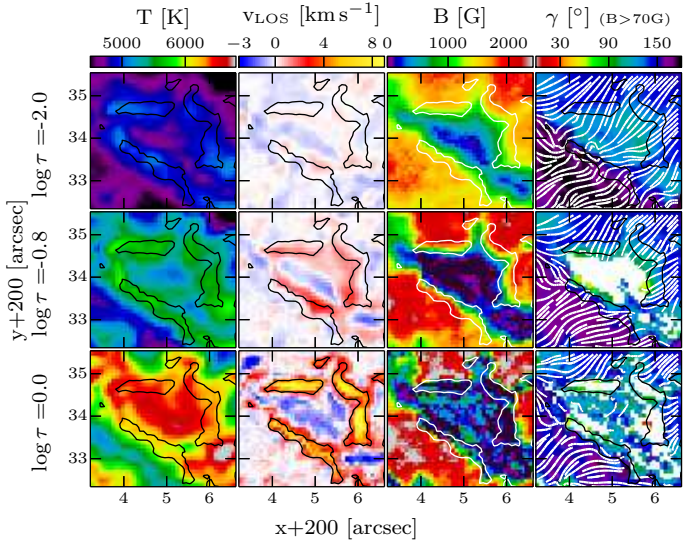
\colfig{lb_granule.noazi}{Light bridge granule (\lbg{}, inset \bx{1} in \fig{icont_vlos}): Plotted are from left to right: temperature $T$, \los{} velocity, magnetic field strength $B$, and inclination $\gamma$ (color scale) of the magnetic field to the \los{} for the three $\log\tau$ nodes used in the inversion (from top to bottom). The azimuth is overplotted as white lines on the inclination plots (rightmost panels). Inclination and azimuth lines are only plotted for $B>70$~G. The black contour lines enclose regions of \los{} velocities greater than +3~\kms{} in the deepest layer.}

\colfig{plg_granule.noazi}{Same as \fig{lb_granule.noazi} for a plage granule (\plg{}, inset \bx{2} in \fig{icont_vlos}). The same color scales were chosen as in \fig{lb_granule.noazi} to facilitate comparison.}
    
\colfig{tqs_granule.noazi}{Same as \fig{lb_granule.noazi} for a quiet-Sun granule (\qsg{}, inset \bx{3} in \fig{icont_vlos}).}

The temperature and the \los{} velocity maps of the \lbg{} (first two columns in \fig{lb_granule.noazi}) show clear evidence for convection: a central upflow of hot material in the deepest layer (bottom row) is surrounded by a downflow of cooler material. With increasing height the granular interior cools down more rapidly than the surrounding, giving the impression of reversed granulation in the highest layer \cite[e.g.,][]{cheung:07b}. The magnitude of both, up- and downflow velocities decreases with height. This general pattern is very similar to that of ``normal'' granules, located in plage and quiet-Sun regions (\plg{}s and \qsg{}s), presented in \fig{plg_granule.noazi} and \fig{tqs_granule.noazi}, respectively.
   
We note that the values of the physical parameters for the ``normal'' granules in plage and quiet-Sun regions (\plg{}s and \qsg{}s) are very similar and show the typical signatures of convective cells. Temperature and \vlos{} stratification in \plg{}s and \qsg{}s are almost indistinguishable and show the expected properties of a central upflow, decreasing in magnitude with height, surrounded by downflows in the intergranular lanes, with the highest velocities in the deepest layer. The magnetic properties of \plg{}s and \qsg{}s differ only in two aspects: The boundaries of \plg{}s show magnetic field strengths that are significantly increased, typically by a few hundred Gauss at the resolution of these data, and the interior of \plg{}s shows an enhanced horizontal magnetic field in the top layer. In contrast, the \qsg{} shows a field-free interior at all heights and only weak fields of on average 100~G in the deepest layer of the intergranular lane. The \plg{}s do appear to be smaller, on average, than the \qsg{}s.

In spite of the general similarity, there are significant differences between the \lbg{}s, on the one hand, and the \qsg{}s and \plg{}s, on the other. Thus, there is a basic difference in the geometry. While in the latter the downflows are localized at roughly the same place at all heights, they lie closer together in the \lbg{}s at higher layers, i.e., the \lbg{}s appear to shrink with height. This difference in geometry was taken into account when making a quantitative comparison between the three different types of granules (\lbg{}s, \plg{}s, and \qsg{}s), e.g., via the averaged numbers summarized in \tab{avgpar}. The values in the table represent mean values of atmospheric parameters over manually selected interiors and boundary regions of five \lbg{}s, \plg{}s, and \qsg{}s each, marked with the red, cyan, and yellow crosses in \fig{icont_vlos}. The interiors of granules are defined as the regions where all three height layers show upflows, in order to reflect the shrinking of especially the \lbg{}s with height. The boundary regions are defined using the fast downflow regions in the deepest layer.

\begin{table}
  \caption[]{Average atmospheric parameters for the interiors and the boundaries of several \lbg{}s, \plg{}s, and \qsg{}s.}
  \label{avgpar}
  \setlength{\tabcolsep}{1.0ex}
  \begin{minipage}{\linewidth}
    \renewcommand{\thefootnote}{\alph{footnote}}
    \centering                          
    \begin{tabular}{lrrrrcrrr}        
    \toprule                 
    \multicolumn{2}{l}{Parameter}& \multicolumn{3}{c}{interior} && \multicolumn{3}{c}{boundary}\\
       \cmidrule{3-5} \cmidrule{7-9}
                  &$\log\tau$& \lbg{}  & \plg{}   & \qsg{}   && \lbg{}   &  \plg{} &  \qsg{}   \\ \toprule
     T            & -2.0    &    4810&   4950&   4870&&  4980 &  5010&   4910 \\
     $[$K$]$      & -0.8    &    5330&   5430&   5360&&  5550 &  5440&   5300 \\
                  &  0.0    &    6590&   6810&   6780&&  6290 &  6340&   6300 \\ \midrule
     B            & -2.0    &     280&    180&     20&&  1400 &   270&     20 \\
     $[$G$]$      & -0.8    &      60&     40&     20&&  1320 &   230&     40 \\
                  &  0.0    &     170&     40&     30&&   320 &   280&     90 \\ \midrule
     $\gamma$     & -2.0    &     120&     92&     89&&   142 &   103&     89 \\
     $[$\grad{}$]$& -0.8    &     108&     94&     91&&   129 &   108&     92 \\
                  &  0.0    &     108&     98&     92&&    95 &   112&     94 \\ \midrule
     \vlos{}      & -2.0    &    -320&   -300&   -250&&   260 &   640&    530 \\
     $[$\ms{}$]$  & -0.8    &    -470&   -730&   -690&&   870 &  1170&    870 \\
                  &  0.0    &    -930&  -1080&   -980&&  5220 &  2210&   1790
 \\ \bottomrule
\end{tabular}
\end{minipage}
\end{table}

The most striking difference in \tab{avgpar} is the downflow velocity in the lowest layer ($\log\tau=0$). These averaged downflows in \lbg{}s are more than a factor of 2 larger than those in the ``normal'' granules. The same is true for the maximum downflow velocities at the boundaries of the \lbg{}s, which reach values of 10~\kms{} compared to the maximum values of 3--4~\kms{} in the darkest parts of the intergranular lanes of ``normal'' granules. The average temperature in the \lbg{}s at the location of these fast downflows ($\log\tau=0$) is similar to the average temperatures at the boundaries of \plg{}s and \qsg{}s, but is on average 140--250~K higher in the middle layers (at $\log\tau=-0.8$). The temperature in the center of the \lbg{} is lower than in the \plg{}, with an increasing difference of $\approx$140~K to $\approx$220~K from the top to the bottom layer. This leads to the interesting feature that the radial temperature profile in the deepest layer of \lbg{}s, i.e. the temperature difference between the granular interior and the boundaries, is rather flat ($\approx$300~K difference) whereas in the ``normal'' granules the interior is significantly hotter than the intergranular lanes. The central upflow above the \lbg{} narrows with height to form a thin, elongated sheet parallel to the light bridge axis. In contrast, the shape of the upflow region in \plg{}s and \qsg{}s remains unchanged with height. The upflow velocities in the interiors at all heights are comparable, reaching maximum values of $\approx$2~\kms{} in the deepest layer and 1~\kms{} in the top layer. 

A striking common feature in the magnetic field strength maps (3$^{rd}$ column of \figs{lb_granule.noazi}--\ref{tqs_granule.noazi}) is the almost complete absence of a magnetic field in the middle and bottom layer of the cellular interior in all three types of granules. For the \lbg{}, a sharp transition exactly at the outer edge of the fast downflows separates these field-free regions from the umbral field with strengths of more than 2~kG at $\log\tau=0$. The top layer of the \lbg{} already exhibits field strengths close to one kG, with lower field strengths of $\approx$200~G within the elongated upflow region. In combination with the magnetic field orientation presented in the inclination maps (4$^{th}$ column of \fig{lb_granule.noazi}) the picture of a cusp-like field configuration closing above the field-free region in the middle to upper photosphere becomes evident. The dark lines overlying the light bridge harboring this \lbg{} in the \caiih{} movie provides further support for this picture. The temperature is lower at the upflow at $\log\tau=-2$ in the \lbg{}, whereas it is relatively flat in the other granules. This is consistent with the presence of a cusp in the field above the upflow lane, so that we see higher and cooler layers there.

A hint of opposite polarity field is present in the inclination maps in the middle and deepest layers of \lbg{}s, located at the innermost boundary of the fast downflows. These opposite polarity fields are very weak. It is therefore difficult to judge how trustworthy their presence is. However, it should be noted that these weak, opposite polarity fields are present at the same location in all \lbg{}s investigated in the course of this study, irrespective of their orientation.

The magnetic field in the plage and the quiet-Sun is concentrated in the intergranular lanes, where it reaches kG and hG values \cite[]{lagg:10b}, respectively (see \figs{plg_granule.noazi} and \ref{tqs_granule.noazi}). In the highest layer the field strength in the \plg{} above the field-free granular interior is slightly above the detection threshold of $\approx$50~G (for horizontal fields). There the field is mainly horizontal and shows a clear preferred orientation toward the spot. This configuration is consistent with a low lying sunspot canopy field extending radially away from the sunspot outside the visible sunspot boundary \cite[]{giovanelli:80,giovanelli:82,solanki:92b,solanki:94,solanki:99a,buehler:14a}. In contrast, the interior of \qsg{}s is field-free at all heights.

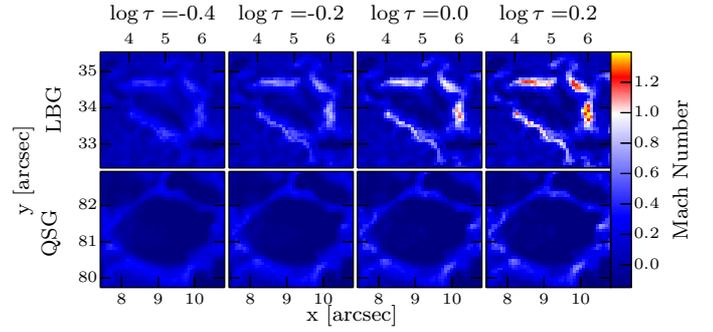
\colfig{machnumber}{Mach number $M$ determined from the the \los{} velocity and the thermodynamic parameters retrieved from the inversion for the \lbg{} (top row, see \fig{lb_granule.noazi}) and a \qsg{} (bottom row, see \fig{tqs_granule.noazi}). The maps only show the very deep photospheric layers from $\log\tau=-0.4$ (left) to $\log\tau=+0.2$ (right). Supersonic velocities are indicated by red/yellow colors, subsonic velocities by blue colors. Negative values denote the Mach number for upflows.}

The fast downflows of up to 10~\kms{} at the edges of \lbg{}s in the deepest layers suggest that these flows are supersonic. To verify this we compute the Mach number $M=v/c_s$ using the thermodynamic parameters gas pressure, $p$, and density, $\rho$, provided by the inversion for every pixel and height grid point in our maps. The equation of state look-up tables from the radiative magneto-hydrodynamic (MHD) code MURaM \cite[]{voegler:05a} are used to compute the sound speed in the general form 
$c_s=\sqrt{\left( \partial p / \partial \rho \right)_{S}}$,
where the partial derivative is taken adiabatically, i.e. at constant entropy $S$. The MURaM tables take into account the effects of partial ionization.
We use \vlos{} to compute a lower limit of the Mach numbers presented in \fig{machnumber} for the \lbg{} (top row, box \bx{1} in \fig{icont_vlos}) and the \qsg{} (bottom row, box \bx{3}). Since the very high speed downflows are concentrated only in the deepest layers of the photosphere, the Mach number maps are only shown for $\log\tau=[-0.4, -0.2, 0.0, +0.2]$ (from left to right). The maps nicely illustrate that in the deepest layers ($\log\tau>-0.2$) the maximum vertical velocities at the edges of the \lbg{} exceed the local sound speed (up to $M\approx 1.5$), whereas in the quiet-Sun the velocities clearly remain subsonic at all heights ($M\lesssim 0.6$).


\subsection{Vertical cuts\label{cuts}}

For the solution of the radiative transfer equation the inversion code SPINOR computes the atmospheric stratification on a much finer vertical (i.e. $\log\tau$) grid than given by the three height nodes at which the atmospheric parameters are varied as part of the $\chi^2$ minimization process. This fine grid allows for vertical cuts through the atmospheres which nicely illustrate the atmospheric stratification in the three different types of granules described in \sect{maps}. It would be preferable to study the height stratification on a geometric scale and not on a $\log\tau$ scale. Unfortunately, the conversion between these scales is not straightforward and requires, among other things, a knowledge of the magnetic curvature force \cite[see, e.g.,][]{mathew:04}, which we do not know how to obtain from the existing SOT/SP data set for the complex field geometries analyzed in this paper. The atmospheric parameters on this finer grid are obtained by a spline interpolation between the three height nodes (for $T$, $\vec{B}$ and \vlos{}). Such cuts are presented in \fig{lb_granule_cut}, \ref{plg_granule_cut} and \ref{tqs_granule_cut} for a \lbg{}, \plg{}, and a \qsg{}, respectively. The locations of these cuts are indicated by the lines marked with \cut{1}, \cut{2}, and \cut{3} in \fig{icont_vlos}. The $\log\tau$ range for these cuts is limited to values between $+0.1$ and $-2.7$, a region where the response functions of the \fei{} 630~nm line pair are sufficiently high to obtain trustworthy values for the atmospheric parameters.

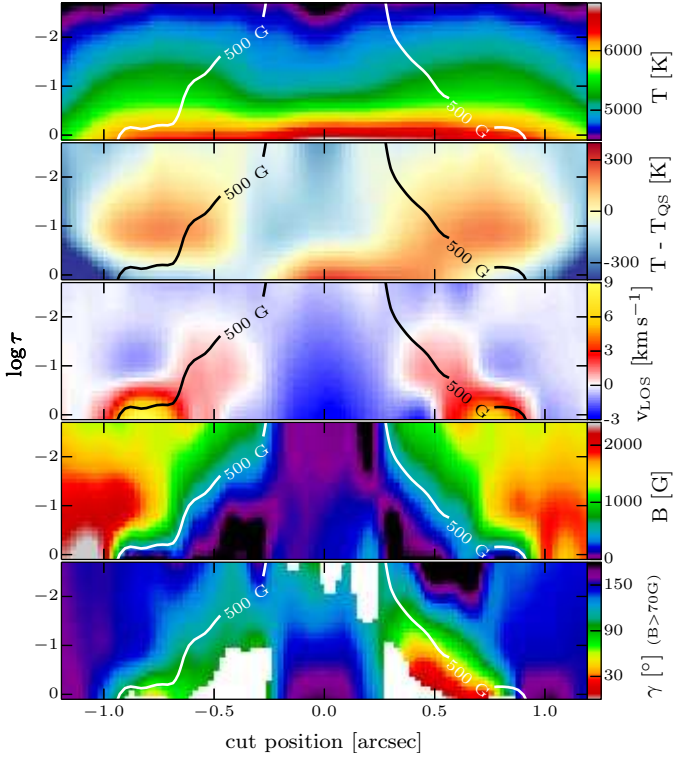
\colfig{lb_granule_cut}{Vertical cut through a \lbg{} (see cut \cut{1} in \fig{icont_vlos}). From top to bottom (panels 1 to 5): temperature, temperature difference to an average quiet-Sun atmosphere, \los{} velocity, magnetic field strength, and magnetic field inclination. The inclination is only shown for field strengths greater than 70~G. A 500~G contour line is overplotted.}

\colfig{plg_granule_cut}{Same as \fig{lb_granule_cut} for a \plg{} (see cut \cut{2} in \fig{icont_vlos}).}

\colfig{tqs_granule_cut}{Same as \fig{lb_granule_cut} for a \qsg{} (see cut \cut{3} in \fig{icont_vlos}).}

The second panel of \fig{lb_granule_cut} emphasizes the enhanced temperature located vertically above the strong downflows, by showing the difference between the temperature across a \lbg{} and an averaged quiet-Sun temperature profile, determined by averaging the temperature profiles over an 8\arcsec$\times$8\arcsec{} large region in the lower lefthand corner in \fig{icont_vlos}. These regions of enhanced temperature difference in \fig{lb_granule_cut} are accompanied by upflows of $\approx$1~\kms{} directly above the high-speed downflows in the deepest layer (see panel 3). The central upflow and the cusp-like shape over the field-free interior of the \lbg{} is nicely outlined in the velocity map in panel 3 and the magnetic field strength map in the fourth panel. The weak opposite polarity fields at the inner edges of the downflows are clearly visible in the inclination map (bottom panel).

\fig{plg_granule_cut} and \ref{tqs_granule_cut} show a vertical cut through a \plg{} and a \qsg{} (cut \cut{2} and \cut{3} in \fig{icont_vlos}) each. Similar to the \lbg{}, the central upflow, surrounded by the downflows in the intergranular lanes, nicely outlines the convective motion in the granule. In contrast to the \lbg{}, the temperature at the boundary of the granule (the intergranular lane) is lower than the averaged quiet-Sun temperature profile. The magnetic field of the \plg{} is concentrated in the highest layer, where it shows a horizontal, low-lying canopy configuration.


\section{Discussion\label{discussion}}

It was shown in \sect{result} that the atmospheric conditions in the lower layers of the \lbg{} interior are qualitatively similar to the interior of ``normal'' granules in plage and quiet-Sun regions. This suggests a common origin for the convective motion creating these cells, which for the granulation pattern in the quiet-Sun is known to be a result of the cooling and hydrogen recombination within the plasma parcels when they reach the low opacity layer of the solar surface. Unlike the surface convection responsible for umbral dots and faint light bridges, which takes place in a 1--2~Mm thick layer immediately below the solar surface \cite[e.g.,][]{schuessler:06}, the convection cells responsible for granular light bridges are rooted deeper in the underlying convection zone \cite[]{rempel:11a}. This interpretation is supported by the near absence of magnetic field in the interior of the \lbg{}s \cite[see also][]{sobotka:13}, distinctly different from the significant field strengths measured in the deep layers of umbral dots, which is lower by only $\approx$500~G than in the surrounding umbra \cite[]{riethmueller:13a}. These field-free regions are only found in the interiors of granular LBs. Narrower LBs however, harbor hecto-Gauss fields in their interior \cite[]{jurcak:06}.

Another indication for the deep anchoring of granular light bridges is their long-term stability. SOT/BFI observations demonstrate, that the granular light bridges in the active region studied here appeared already 4--5 days before the presented SOT/SP scan and lasted another 4--5 days after this scan, until the sunspot finally started to decay (see the animation attached to \fig{gband_gap}). Granular light bridges are therefore likely to be real gaps in the subsurface magnetic environment. According to \cite{rempel:11a}, such gaps may be the result of fragmentation events, where field-free plasma intrudes the magnetic root of the sunspot several Mm below the solar surface. In his magneto-hydrodynamic simulations, such intrusions become visible in the photosphere after timescales of several hours to one day. An alternative possibility is that \lbg{}s are found at the boundary between fragments that emerged individually and then joined to form the sunspot \cite[]{garciadelarosa:87}.

The differences between \lbg{}s and ``normal'' granules are clearly dominated by the special location where the upward moving gas reaches the photosphere. The \lbg{}s are exposed to the cold, strongly Wilson-depressed umbral environment, where they stick out like a few-hundred-kilometer-high mountain ridge crossing the umbra \cite[]{ruedi:95c,lites:04}. The downflow velocities of up to 10~\kms{} at the boundaries of this ridge can be explained by a combination of gravitational acceleration and the efficient radiative cooling toward the cold umbral environment. This cooling causes the gas to sink faster, in effect making the convection more vigorous. Another reason for the strong observed downflows in \lbg{}s is that we see these downflows at geometrical heights significantly deeper than the downflows in normal intergranular lanes. This is caused by the inclined walls of the umbra as the magnetic field expands with height (and decreasing gas pressure), directly visible in the vertical cut plotted in \fig{lb_granule_cut}. Since granular downflows tend to accelerate with depth, at least for the first few 100~km, this view into deeper layers also tends to show stronger downflows. As already stated by \cite{shimizu:11}, a reconnection mechanism, as proposed by \cite{louis:09}, is not required to produce these downflows.

\fig{machnumber} demonstrates that the gas flows in the deepest layer of the edges of \lbg{}s are supersonic. The possible existence of transonic fluid velocities was postulated by \cite{cattaneo:90} and \cite{malagoli:90} using numerical simulations of convection. An important ingredient to these flows are non-adiabatic effects caused by the radiative losses in this layer, causing the temperature and subsequently the sound speed to decrease significantly and therefore to accelerate the pressure gradient driven, initially horizontal flow to transonic speeds near the edges of granules. Such horizontal supersonic flows have been detected in observations 
\cite[]{solanki:96a,rybak:04,bellotrubio:09}, but this is the first time that granulation with supersonic downflows is reported. In the case of the elevated \lbg{} discussed here, the radiative losses do not act on the horizontal flows above the granule interior, but continue to be an efficient, non-adiabatic process lowering the temperature as the gas flows down the slanted walls of the \lbg{}. This process, in combination with the gravitational acceleration, provides a natural explanation for the supersonic flow speeds observed at the edges of \lbg{}s.

The temperature of the downflowing material is determined by multiple effects. On the one hand, it is lowered by radiative cooling, both into space and into the neighboring cold umbra. On the other hand, the deeper layers are hotter due to the general increase of temperature with depth. The balance between these two effects results in temperatures of the visible downflowing gas that are rather similar to the upflowing gas in the lower photosphere (see \fig{lb_granule.noazi}), in contrast to a \plg{} or \qsg{} (see \fig{plg_granule.noazi} and \ref{tqs_granule.noazi}).


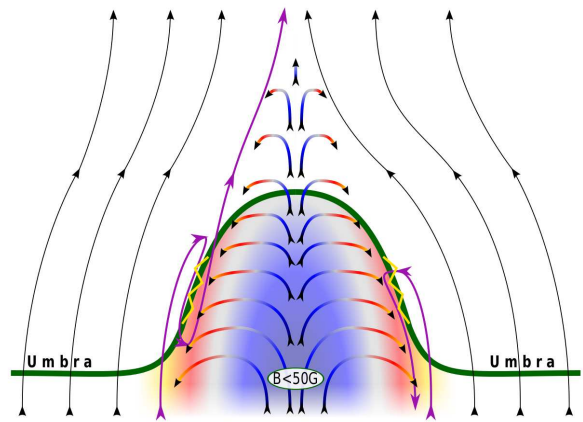
\colfig{lb_granule_fielddrag}{Sketch of the magnetic and velocity field configuration in a light bridge granule. The solar surface is indicated by the thick green line. The upflowing material in the nearly field-free cell interior is depicted by the blue parts of the curved arrows. The downflowing material (red lines) is able to drag down magnetic field lines (black and purple), creating a region where additional heating might occur (yellow crinkled line). The background colors in the cell interior indicate the \los{} velocities (upflows: blue, downflows: red/yellow).}

The observed downflows occur in a regime where the kinetic energy dominates over the magnetic energy, i.e., where the magnetic field strength is below the equipartition field strength ($B_{\mbox{eq}}=v\sqrt{\mu_0 \rho}$, with $v$ being the typical velocity of motion and $\mu_0$ the magnetic permeability). As a consequence, the downflowing material is able to drag the outermost umbral magnetic field lines down and bend them back into the solar interior. This scenario is illustrated in the sketch in \fig{lb_granule_fielddrag}. The magnetic field configuration determined from the inversion is compatible with two different scenarios: The tension of the magnetic field is high enough to allow the field line to reverse its direction again (lefthand side of \fig{lb_granule_fielddrag}), or the field is dragged down and eventually probably ``shredded'' in the convective motions in the granule interior (right side of \fig{lb_granule_fielddrag}). The opposite polarity field measured at the inner edge of the fast downflows provides evidence for these scenarios. In both cases, the opposite polarity field, confined to a narrow layer, may dissipate a part of its energy, indicated by the yellow zig-zag line in \fig{lb_granule_fielddrag}. Magnetic energy can be released by either reconnection processes or by Ohmic dissipation of electric currents flowing in these narrow layers.

The apparent temperature enhancement directly above the fast downflow regions can be attributed to such a magnetically driven heating mechanisms only to a minor extent. As shown in \oapp{deltat} one needs to dissipate $\approx$500~G in order to raise the temperature by $\approx$100~K at $\tau \approx 1$. Due to radiative losses, this enhancement is soon removed, so that very significant amounts of magnetic flux would have to be constantly removed to achieve any measurable heating. The often observed enhanced chromospheric activity above light bridges in the form of jets and surges \cite[e.g.,][]{shimizu:09,shimizu:11,bharti:07} may, however, be the result of reconnection triggered by the reversal of the field caused by the downflowing material.
Since no enhanced chromospheric activity was observed during the time of the SOT/SP scan, there is little evidence for such a magnetically driven heating mechanism.

The transonic speeds of these downflows must unavoidably lead to the formation of shocks, when the flows encounter the high density, deep photospheric layers. The resulting shock waves could, outside the downflow channel, in principle propagate upward into the umbra and subsequently heat the layers above the downflows, explaining the observed apparent temperature enhancement. However, since the measured downflow speeds continue to increase with decreasing optical depth, the shock must be located in deeper layers not accessible by our observations. Therefore the energy deposited by this process is unlikely to reach the heights where we observe the apparent temperature enhancement.

Two other possible origins for this apparent temperature enhancement do not require a specific heating mechanism: Firstly, this enhancement could result from the energy radiated horizontally away from the, on a geometrical height scale elevated, slanted walls of the \lbg{} into the umbra. Due to the decreasing size of the \lbg{} with height the regions heated by this energy appear directly above the strong downflows in the deepest layer, where the granule is broadest. Secondly, the magnetic field in the downflow lanes of the \lbg{} is stronger than in the other two granules studied here and in addition is inclined above the \lbg{} (cusp shape), which decreases the effective gravity and increases the vertical scale height. The density is reduced by the magnetic field, resulting in a depression of the iso-$\tau$ surfaces to deeper, hotter layers. The increased vertical scale height along the field reduces the vertical temperature gradient, producing an apparent temperature enhancement as compared to the surrounding areas with weaker and more vertically oriented magnetic field. Since this effect is mainly produced by the height variation of the iso-$\tau$ surfaces, it may well be absent if geometric height coordinates would be used.
However, the lack of knowledge about the true geometric height scale makes it difficult to estimate the significance of the above mentioned processes in producing this apparent temperature enhancement.


At higher layers above the \lbg{}s, the expected cusp-like configuration of the magnetic field becomes clearly visible in the inversion results. The field reaches inclinations with respect to the umbral field of 70$^\circ{}$, in good agreement with the value found by \cite{scharmer:08a} on a short, irregular light bridge. A narrow central upflow lane remains visible up to the $\log\tau=-2.5$, i.e. the highest level reliably retrieved by inversions of the \fei{} 630~nm line pair. Since no continuous net upflow above light bridges is observed at chromospheric heights \cite[]{joshi:14a}, it is likely that at heights above the formation height of these \fei{} lines the upflowing material reverses its direction and contributes to the observed downflows.

The visual impression from the velocity maps in \fig{lb_granule.noazi} suggests a significant excess of downflowing material over upflowing material. It is likely that this impression is a consequence of the fact that the up- and downflowing material is measured at different heights, with the downflowing material being sampled at deeper layers, and, because of the inclined iso-$\tau$ surfaces, over a range of heights. The corrugation of the $\log\tau=0$ surface makes it virtually impossible to establish an overall mass flux balance. This problem may be solved in the future by stereoscopic measurements, e.g., by combining magnetic field maps obtained with the Solar Orbiter Polarimetric and Helioseismic Imager \cite[]{gandorfer:11a} with ground-based or Earth-orbiting spectropolarimetric measurements.

\section{Summary \& Conclusion\label{conclusion}}

We presented results from spatially coupled inversions of Stokes profiles in granular light bridges and in plage and quiet-Sun granules. A significant degree of similarity between light bridge granules (\lbg{}s) and granules in plage and quiet-Sun regions (\plg{}s and \qsg{}s), especially in the deep layers of the cell interior, point to the common driving mechanism of the convective motions. The interiors of all three types of granules are void of measurable magnetic field in their deepest observable layers ($\tau\simeq 1$). The field-free regions are dominated by upflowing plasma with velocities of up to 2~\kms{}. For \lbg{}s, these upflows get squeezed in higher layers into narrow, thin sheets by the expanding magnetic field of the umbra on both sides of the \lbg{}s. The magnetic configuration is consistent with a cusp overlying the upflow. The walls of the \lbg{}s, exposed to the dark umbral environment, harbor downflows with velocities of up to 10~\kms{}, exceeding the local sound speed in the deepest observable layers. Hints of field reversal are present in the vicinity of these downflows.


The similarity between \lbg{}s and ``normal'' granules suggests that granular light bridges are anchored in deep layers. This distinguishes granular light bridges from other convective processes in sunspot umbrae, like umbral dots or faint light bridges, which are, according to MHD simulations, the product of surface magneto-convection within the 1--2~Mm just below the local solar surface.

The exposure of the walls of granular light bridges due to reduced opacity in sunspot umbrae offers an attractive way to probe the deep interior of convective cells using \lbg{}s. A future analysis of \lbg{}s under different viewing geometries, either by studying their center-to-limb variation or by performing stereoscopic measurements might help to uncover further details of magnetoconvection, such as the confirmation of the presence of heating zones, or the study of flow geometry and magnetic field configuration on a geometrical height scale.

\begin{acknowledgements}
Hinode is a Japanese mission developed and was launched by ISAS/JAXA, collaborating with NAOJ as a domestic partner, NASA and STFC (UK) as international partners. Scientific operation of the Hinode mission is conducted by the Hinode science team organized at ISAS/JAXA. This team mainly consists of scientists from institutes in the partner countries. Support for the post-launch operation is provided by JAXA and NAOJ (Japan), STFC (U.K.), NASA, ESA, and NSC (Norway). This work was partly supported by the BK 21 plus program through the National Research Foundation (NSF) funded by the Ministry of Education of Korea. The development of the inversion code benefited from two meetings held in February 2010 and December 2012 at the International 
Space Science Institute (ISSI) in Bern (Switzerland) as part of the International Working Group {\it ``Extracting Information from Spectropolarimetric Observations: Comparison of Inversion Codes''}.
\end{acknowledgements}


\Online

\begin{appendix}

\section{Temperature increase due to magnetic field dissipation \label{deltat}}

As stated in \sect{discussion}, the enhanced temperature observed directly above the regions of fast downflows at the edges of \lbg{}s can only to a minor extent be attributed to magnetically driven heating mechanisms. To demonstrate this, we compute here the maximum possible temperature increase ($\Delta T$) by completely dissipating a magnetic field ($B$) with a given magnetic energy density ($\rho_M=B^2/2\mu_0$) under typical photospheric conditions:
\begin{equation}
\label{dt}
\Delta T = \frac{Q}{c} = \frac{Q}{c_{\mbox{mol}} n}
\end{equation}
with $Q$ being the thermal energy, $c$ the heat capacity, $n$ the  amount of gas in moles, and $V$ the volume. For simplicity we assume a 1-atomic, ideal gas ($c_{\mbox{mol}} = \frac{3}{2} R$) to compute $Q$:
\begin{equation}
Q = V \rho_M = V \frac{B^2}{2 \mu_0} = \frac{n R T}{p}\frac{B^2}{2 \mu_0}
\end{equation}
(with $R$ = universal gas constant, $p$ = gas pressure, and $T$ = temperature). By inserting this into \eq{dt} we can compute $\Delta T$:
\begin{equation} 
\Delta T = \frac{2}{3} \frac{T}{p} \frac{B^2}{2 \mu_0}
\end{equation}
Using typical atmospheric conditions in a sunspot from the umbral model of \cite{maltby:86} at $\tau=1$ ($p=2\cdot 10^5$~dyn/cm$^2$, $T=3500$~K) we obtain a temperature increase of $\Delta T = 18.5$~K for $B=200$~G, and $\Delta T = 116$~K for $B=500$~G.

\section{Animations}

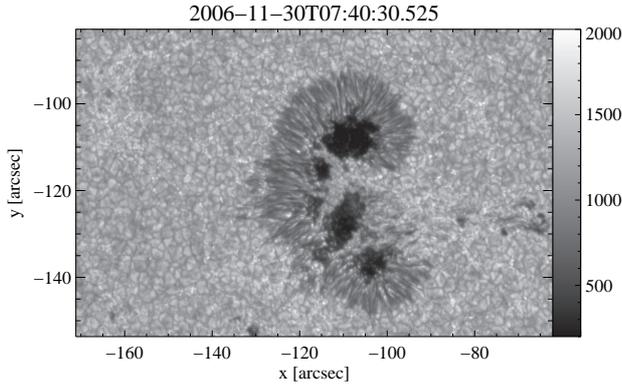
\colfig{gband_gap}{G-band images demonstrating the long-term stability of the \lbg{}s under investigation in this paper. The animation, composed from G-band images of the  Hinode SOT Broad-band Filter Imager (BFI), covers the time period from 2006-Nov-30, 07:40~UT until 2006-Dec-03, 23:59~~UT. The same granular light bridges are present from the beginning of the observations until the end. The movie is also available on the MPS website:  \href{http://www.mps.mpg.de/homes/lagg/OnlineMaterial/2014_LightBridge/gband_gap.avi}{http://www.mps.mpg.de/homes/lagg/OnlineMaterial/}
\href{http://www.mps.mpg.de/homes/lagg/OnlineMaterial/2014_LightBridge/gband_gap.avi}{2014\_LightBridge/gband\_gap.avi}.}

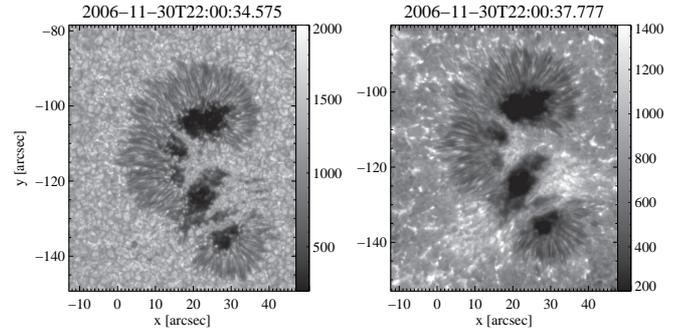
\colfig{ca_gb}{BFI images of the G-band (left) and in the \caiih{} line (right) around the time of the SOT/SP scan discussed in the paper (\fig{icont_vlos}). In the animation, the exact time of the SOT/SP scan is indicated by the red text label in the upper lefthand corner. The movie demonstrates the absence of enhanced chromospheric activity above the \lbg{}s during the time of the SOT/SP scan. The movie is also available on the MPS website: \href{http://www.mps.mpg.de/homes/lagg/OnlineMaterial/2014_LightBridge/ca_gb.avi}{http://www.mps.mpg.de/homes/lagg/OnlineMaterial/}
\href{http://www.mps.mpg.de/homes/lagg/OnlineMaterial/2014_LightBridge/ca_gb.avi}{2014\_LightBridge/ca\_gb.avi}.}

\end{appendix}

\end{document}